\documentclass[twocolumn,showpacs,preprintnumbers,amsmath,amssymb,prl]{revtex4}
\usepackage{pifont}
\usepackage{amsmath}
\usepackage{amsfonts}
\usepackage{graphicx}
\usepackage{amssymb}
\usepackage{dcolumn}
\usepackage{bm}
\usepackage{slashed}

\begin{document}

\title{Complete QED theory of multiphoton trident pair production in strong laser fields}

\author{Huayu Hu$^{1,2}$}
\author{Carsten M\"uller$^1$}
\email[Corresponding author: ]{c.mueller@mpi-k.de}
\author{Christoph H. Keitel$^1$}
\affiliation{$^1$Max-Planck-Institut f\"ur Kernphysik, Saupfercheckweg 1, 69117 Heidelberg, Germany\\
$^2$Department of Physics, National University of Defense Technology, Changsha 410073, P. R. China}

\date{\today}

\begin{abstract}
Electron-positron pair creation by multiphoton absorption in the
collision of a relativistic electron with a strong laser beam is
calculated within laser-dressed quantum
electrodynamics. Total production rates, positron spectra, and relative
contributions of different reaction channels are obtained in various
interaction regimes. We study the process in a manifestly
nonperturbative domain which is shown accessible to future experiments 
utilizing the electron beamlines at novel x-ray laser facilities or 
all-optical setups based on laser acceleration. Our theory moreover allows us
to add further insights into the experimental data from SLAC [D.\,Burke
{\it et al.}, Phys.\,Rev.\,Lett.\,{\bf 79}, 1626 (1997)].
\end{abstract}


\maketitle

Quantum electrodynamics (QED) describes the interaction of electrons and photons. When the photon source is an intense laser field, electrons may couple nonlinearly to the field giving rise to multiphoton processes. Theoreticians started to consider nonlinear QED phenomena such as multiphoton Compton scattering or $e^+e^-$ pair production soon after the invention of the laser \cite{Reiss,Nikishov}.

In the mid 1990s pioneering studies at SLAC (Stanford, California) revealed nonlinear QED effects in experiment. In particular, the first and so far unique observation of multiphoton pair production was accomplished in collisions of the $\approx50$ GeV electron beam from SLAC's linear accelerator with an intense laser pulse \cite{SLAC}. The laser frequency and field strength were largely Doppler-enhanced in the rest frame of the high-energy projectile.

The SLAC experiment has triggered substantial theoretical efforts on laser-induced $e^+e^-$ pair creation in the last decade (see \cite{Report,Ehlotzky} for reviews). Most studies consider pair creation in laser-proton collisions (e.g., \cite{MVG,Milstein,Kaminski,Kuchiev,Deneke,ADP}) or in counterpropagating laser beams (e.g., \cite{Ringwald,Alkofer,Narozhny,Kirk,Gies,Grobe}). Meanwhile, theoreticians are already considering more refined aspects of the process such as final-state pair correlations \cite{Kuchiev}, the influence of more complex fields \cite{Gies,ADP}, and the creation of $\mu^+\mu^-$ pairs \cite{Kuchiev,Deneke}.

Nevertheless, despite all these efforts, no complete QED
calculation of the SLAC experiment exists as yet. One purpose of the present
study is to fill this gap in the theory of strong-field phenomena, 
taking advantage of the recent theoretical progress and advancing it further. Indeed, nonlinear pair creation, both in the perturbative and nonperturbative domains, is nowadays becoming accessible to all-optical setups involving laser
acceleration devices. Note that a recent experiment on powerful laser-solid interaction observed abundant (linear) pair creation through the conversion of $\gamma$-ray bremsstrahlung in the field of gold nuclei \cite{Chen}.

In comparison with multiphoton pair creation in laser-proton collisions, the theoretical consideration of laser-electron collisions is rendered more involved in several respects. One needs to take into account (i) the dressing of the projectile electron by the laser field, (ii) the recoil the projectile suffers during the collision, (iii) the indistinguishability of the scattered projectile with the created electron of the pair (Pauli principle), and (iv) the possibility of real photon emission by the projectile in the field (Compton scattering).

In laser-electron collisions, two pair creation processes are usually distinguished. The first is of Bethe-Heitler type; the pair is produced by the absorption of $N$ laser photons in the Coulomb field of the incoming electron:
\begin{eqnarray}
\label{trident}
e + N\omega \to e' + e^+e^-.
\end{eqnarray}
The second is a two-step process where first a high-energy $\gamma$-photon is generated
by Compton backscattering off the electron beam,  which afterwards creates the pair in a photon-multiphoton collision \cite{Reiss,Nikishov}:
\begin{eqnarray}
\label{twostep}
\gamma + N\omega \to e^+e^-.
\end{eqnarray}
Reaction (\ref{twostep}) represents the strong-field generalization of the process $2\gamma\to e^+e^-$ first studied by Breit and Wheeler \cite{Breit} and exhibits a nonperturbative nature at very high fields. We name it as multiphoton Breit-Wheeler process.

The analysis of SLAC's experimental data \cite{SLAC} relied on separate simulations of both processes. However, while for the multiphoton Breit-Wheeler reaction a sophisticated nonperturbative theory was available \cite{Reiss,Nikishov}, the contribution from the direct process was estimated in a rather approximate manner \cite{internal1} based on the Weizs\"acker-Williams method since a formal theory did not exist.

In this Letter we provide a nonperturbative laser-dressed QED calculation of multiphoton
trident pair creation in laser-electron collisions. Our approach
treats the competing processes (\ref{trident}) and (\ref{twostep}) in a unified way
and opens deeper insights into the SLAC measurements. Further, we
evaluate the creation rates in the fully nonperturbative regime which could be 
probed by future experiments employing upcoming technologies.

\begin{figure}
\includegraphics[height=2.7cm,width=6cm]{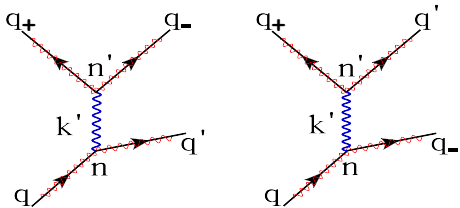}
\caption{\label{feynman} Furry-Feynman diagrams of multiphoton trident pair production in electron-laser collisions. The zigzag-lines represent the exact lepton wave-functions in the laser field (Dirac-Volkov states \cite{Volkov}) and are labeled by the laser-dressed particle momenta. In the left diagram, the incoming electron scatters from a state of dressed momentum $q$ to $q'$ by absorbing $n$ laser photons and emitting an intermediate photon, which afterwards decays into an $e^+e^-$ pair upon absorption of $n'$ laser photons. The corresponding exchange diagram is shown on the right. In the weak-field limit ($\xi\ll 1$), the diagrams can be expanded in Feynman diagrams of field-free QED. For a total number of $N=n+n'=1$ absorbed photons, eight leading-order diagrams arise this way \cite{trident}. The case $N=6$ corresponds to 168 such diagrams.}
\end{figure}

The laser-dressed Feynman diagrams of multiphoton trident pair creation are depicted in Fig.\,\ref{feynman}. The leptons are described by Dirac-Volkov states \cite{Volkov} which fully account for their interaction with the  external plane-wave laser field (Furry picture). To leading order in the QED finestructure constant $\alpha$ the amplitude for the process reads $S_{fi}=\mathcal{M}(q,q',q_+,q_-)-\mathcal{M}(q,q_-,q_+,q')$,
where
\begin{align}\label{smatrix1}
\mathcal{M}(q,q',q_+,q_-)=&-i\alpha\int d^4x\int d^4y \bar{\Psi}_{q'}(x) \gamma_\mu \Psi_{q}(x)\nonumber\\
&\times D^{\mu\nu}(x-y)\bar{\Psi}_{q_-}(y) \gamma_\nu \Psi_{q_+}(y)\,,
\end{align}
and the exchange term $\mathcal{M}(q,q_-,q_+,q')$ correspond to the left and right diagrams in Fig.\,\ref{feynman}, respectively. Here, $\Psi_q, \Psi_{q'}, \Psi_{q_+}, \Psi_{q_-}$ denote the laser-dressed lepton states and $D^{\mu\nu}(x-y)=\int\frac{d^4k'}{(2\pi)^4}\frac{-i g^{\mu\nu}}{k'^2}e^{ik'\cdot (x-y)}$
is the free photon propagator. Relativistic units with $\hbar=c=\epsilon_0=1$ are used. In the presence of the field, the particles are characterized by their dressed four-momentum $q^\mu$, which is related to the free four-momentum $p^\mu$ outside the field by $q^\mu=p^\mu+\frac{m^2 \xi^2}{2 k\cdot p}k^\mu$. Here, $m$ is the electron mass, $k^\mu$ the laser wave four-vector, and $\xi=\frac{e}{m}{\bar A}$, with the root-mean-square value ${\bar A}$ of the laser vector potential.

After Fourier series expansions of the periodic parts in the laser-dressed leptonic transition currents in Eq.\,(\ref{smatrix1}), the space-time integrations can be performed. The amplitude adopts the form
\begin{align}\label{M_fi}
\mathcal{M}(q,q',q_+,q_-)&=\displaystyle\sum_{N\ge N_0}\sum_{n=-\infty}^\infty \frac{\delta^{(4)}(q+Nk-q'-q_+-q_-)}{(q-q'+nk)^2}\nonumber\\
&\times M^\mu(q,q'|n) M_\mu(q_+,q_-|N-n)\, ,
\end{align}
where $N$ is the total number of photons absorbed, $n$ is the number of photons absorbed at the first vertex (see Fig.\,\ref{feynman}), and $M^\mu(q,q'|n)$, $M_\mu(q_+,q_-|N-n)$ are complex functions of the particle momenta and laser parameters. The threshold photon number $N_0$ following from four-momentum conservation is the smallest integer $N$ with
\begin{equation}\label{threshold}
N \omega' \geq 4 m_*\,,
\end{equation}
where $\omega'$ is the laser photon energy in the average rest frame of the projectile and $m_*$ the dressed electron mass defined as $m_*^{2}=q^2=m^2 (1 + \xi^2)$. The intensity dependence of the dressed mass influences the threshold; for example, $N_0=6$ for the SLAC parameters $p^0=46.6$\,GeV, $\omega=2.35$\,eV and $\xi\approx 0.3$ \cite{SLAC}, whereas $N_0=5$ photons would suffice in a weaker field ($\xi < 0.22$).

The total rate is obtained as
\begin{equation}\label{rate}
R = \frac{1}{T} \int\frac{d^3q_+}{(2\pi)^3} \int\frac{d^3q_-}{(2\pi)^3} \int\frac{d^3q'}{(2\pi)^3} \frac{1}{4}\sum_{\rm spins} |S_{fi}|^2\,
\end{equation}
with the interaction time $T$ and a statistical factor $1/4$ due to initial spin averaging and the two identical final-state electrons. The spin sum can be converted in the usual way into voluminous trace products. The multi-dimensional integration in Eq.\,(\ref{rate}) is carried out numerically by an appropriate Monte Carlo routine.

Note that in the weak-field, one-photon limit ($\xi\ll 1$, $N=1$) our approach reproduces the well-known cross section for pair creation by a single $\gamma$-photon on an electron \cite{trident}. The limit $\omega\to 0$ corresponds to trident pair production in constant crossed fields \cite{Ritus}. As a general rule, the rate for an $N$-photon process in a laser field scales as $\xi^{2N}$ in the perturbative regime ($\xi\ll1$).

We emphasize that our laser-dressed QED approach to multiphoton trident pair creation incorporates the direct process as well as the two-step mechanism [see Eqs.\,(\ref{trident}), (\ref{twostep})]. Due to the presence of the laser field, the intermediate photon may reach the mass shell and become real ($k'^2=0$), provided $N\ge 2$ holds. The second-order Feynman graph then decomposes into two first-order diagrams which precisely correspond to the multiphoton Compton and Breit-Wheeler sub-processes \cite{nomenclature}.

For on-shell photon momenta $k'^2=(q-q'+nk)^2=0$, as seen in Eq.\,(\ref{M_fi}), the photon propagator formally diverges which is mainly due to the infinite spatio-temporal extent of the laser field assumed. A regularization procedure transforms the divergences into finite physical resonances. These resonances have been studied most extensively with respect to laser-assisted M{\o}ller scattering \cite{Ehlotzky}. In our case we regularize the propagator by taking the finite interaction time $T$ into account, according to
$1/{k'^2} \to 1/(k'^2+i\varepsilon)$ with $\varepsilon = 2|{k'}^0|/T$. In the SLAC experiment \cite{SLAC}, the electron passes through the laser focus in roughly $T_0\approx 40$\,fs (focal waist size $\sim 4\mu$m, beam crossing angle 17$^\circ$). This (lab-frame) interaction time will be used in the following \cite{Decay}. 

\begin{figure}
\includegraphics[width=7.5cm]{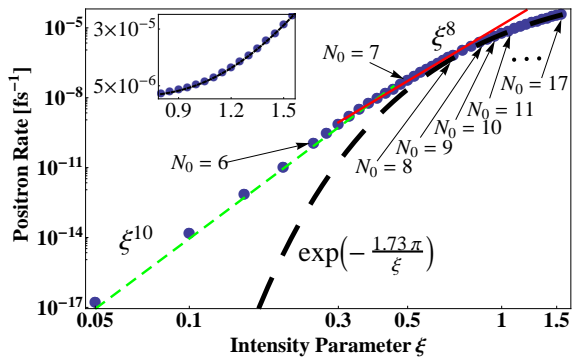}
\vspace{-0.3cm}
\caption{\label{power2} (color online) Positron rate dependence on $\xi$ in the head-on collision of a 46.6\,GeV electron with a linearly polarized 527\,nm laser wave. The blue dots denote our numerical results from Eq.\,(\ref{rate}). The green short-dashed and red solid lines, respectively, are $\xi^{10}$ and $\xi^{8}$ power-law fits to the data, and the black long-dashed line is an exponential fit of the form $\exp[-\theta\pi/\xi]$ with $\theta=1.73$. $N_0=6,\ldots,17$ indicate the minimum numbers of laser photons in different $\xi$ regimes [see Eq.\,(\ref{threshold})]. The inset shows an enlargement of the nonperturbative $\xi\sim 1$ domain on a linear scale.}
\end{figure}

Fig. \ref{power2} shows the trident pair production rate as a function of the laser intensity parameter $\xi$. For the electron momentum and laser frequency chosen, the $\xi$ dependence gradually changes from a $\xi^{10}$ behavior ($\xi\lesssim 0.3$) to a flatter $\xi^8$ increase ($\xi\sim 0.5$), and eventually leads into an exponential dependence ($\xi\sim 1$) similar to the famous Schwinger rate \cite{Report} and marking the transition into the fully nonperturbative regime. Here, photon orders up to $N \approx 50$ give significant contributions to the total rate.

The SLAC experiment found a rate scaling of $R\sim\xi^{10}$ around $\xi\approx 0.3$ \cite{SLAC}, appearing indicative of the typical $\xi^{2N_0}$ dependence in the perturbative domain. However, as mentioned above,  $N_0=6$ in the SLAC case. In fact, our simulation for $\xi=0.3$ reveals that, on average, 6.44 photons are absorbed in total, with 1.62 (4.82) photons being absorbed at the first (second) vertex in Fig.\,\ref{feynman}. Hence, in contrast to the common interpretation, the SLAC experiment did not operate in the perturbative $\xi^{2N_0}$-domain but rather observed the onset of nonperturbative effects which would become more pronounced for $\xi\approx 1$ (see Fig. \ref{power2}). The continuation of the $\xi^{2N_0}$ behavior at $\xi\ll 1$ into the intensity domain where this $N_0$-th order channel has already closed was discussed in \cite{Reiss2} and shown for the example of strong-field atomic ionization.

From Fig.\,\ref{power2} a total lab-frame positron rate per projectile of $R\approx4\times10^4$\,s$^{-1}$ 
results by averaging over a Gaussian laser focus peaked at $\xi=0.3$. It is in reasonable agreement with the experimental result, where about 100 $e^+$ were created in 22,000 shots from 10$^7$ electrons in the interaction region \cite{SLAC}. The corresponding rate is slightly smaller, $R_{\rm exp}\approx10^4$\,s$^{-1}$, which is mainly attributable to shot-to-shot intensity variations ($0.2\lesssim\xi_{\rm peak}\lesssim 0.3$).

\begin{figure}
\includegraphics[width=4.2cm]{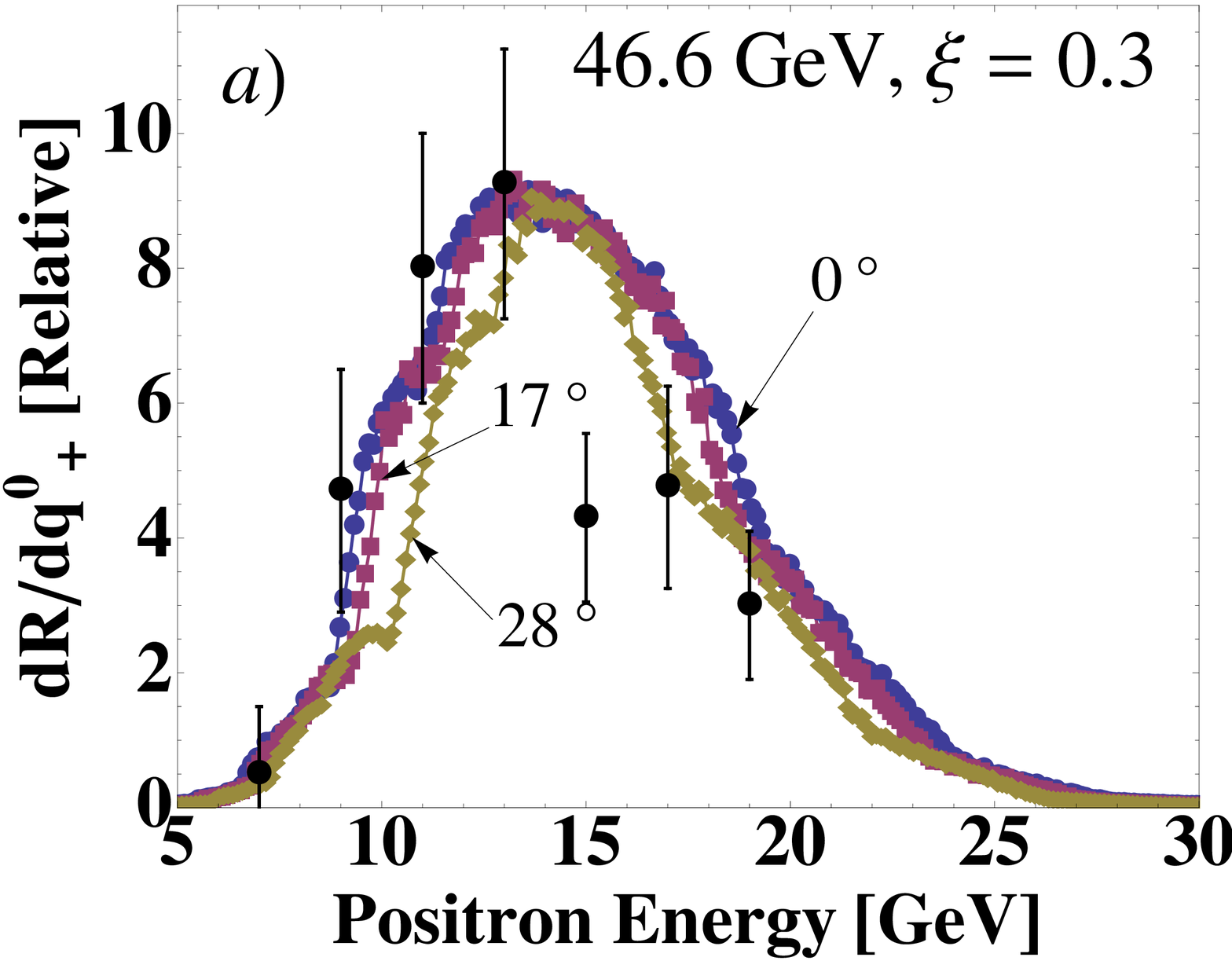}
\includegraphics[width=4.2cm]{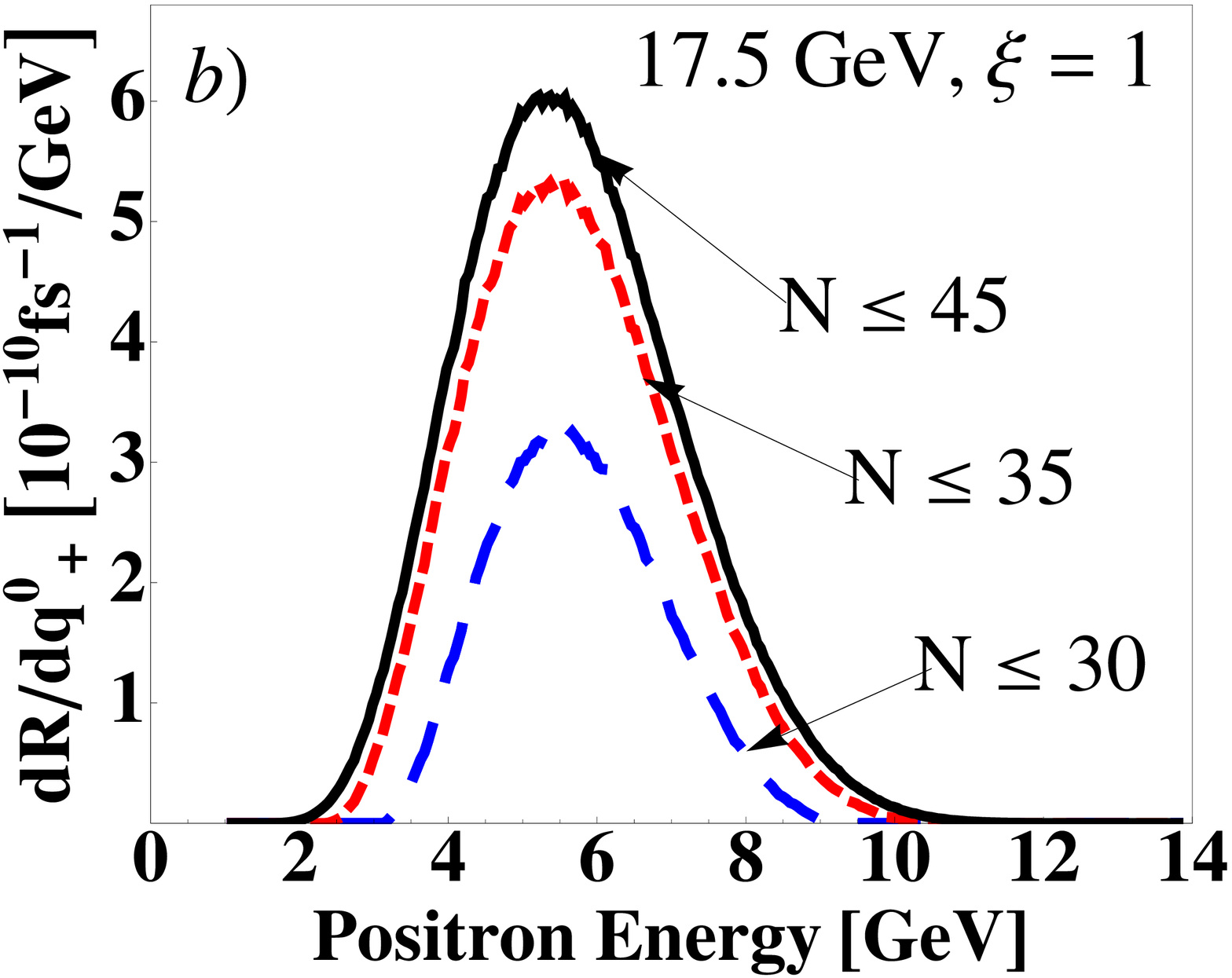}
\vspace{-0.3cm}
\caption{\label{positron} (color online) Positron energy spectra in electron collisions with a 527\,nm laser beam. (a) $\xi=0.3$ and $p^0=46.6$\,GeV; the dots with an error bar are the experimental data from \cite{SLAC}. The three calculated curves refer to different center-of-mass energies, corresponding to different beam crossing angles as indicated, and have been normalized to the same height to facilitate their comparison. (b) Nonperturbative domain at $\xi=1$ and $p^0=17.5$\,GeV; the blue long-dashed (red short-dashed) line refers to the partial spectrum including up to 30 (35) absorbed photons. The black solid line shows the full spectrum ($N\leq 45$).}
\end{figure}

The energy distribution of the positrons is displayed in Fig.\,\ref{positron}. Our calculations for the SLAC parameters \cite{SLAC} reproduce the measurements very well (apart from the data point at $q_0^+\approx 15$\,GeV), see Fig.\,\ref{positron}(a). In addition we find that the spectral maximum remains at $q_0^+\approx 13$\,GeV when the collision angle is varied from $0^\circ$ to $28^\circ$. This explains why the peak position in the experiment is not blurred by averaging over the laser focus and electron beam profile. We note besides, that the mean energy of the recoiled electron after head-on collision is 16.8 GeV, corresponding to an energy loss of $64\%$.

The fully nonperturbative regime could be probed, e.g., by utilizing 17.5\,GeV electrons from the upcoming European XFEL beamline at DESY (Hamburg, Germany) \cite{DESY} combined with a table-top 10\,TW laser system. Here, the contributions from many photon orders form the $e^+$ spectrum, see Fig.\,\ref{positron}(b). Such an experiment would represent a non-standard application of the XFEL electron beam, usually serving to generate x-ray light.

\begin{figure}
\includegraphics[height=3cm,width=7cm]{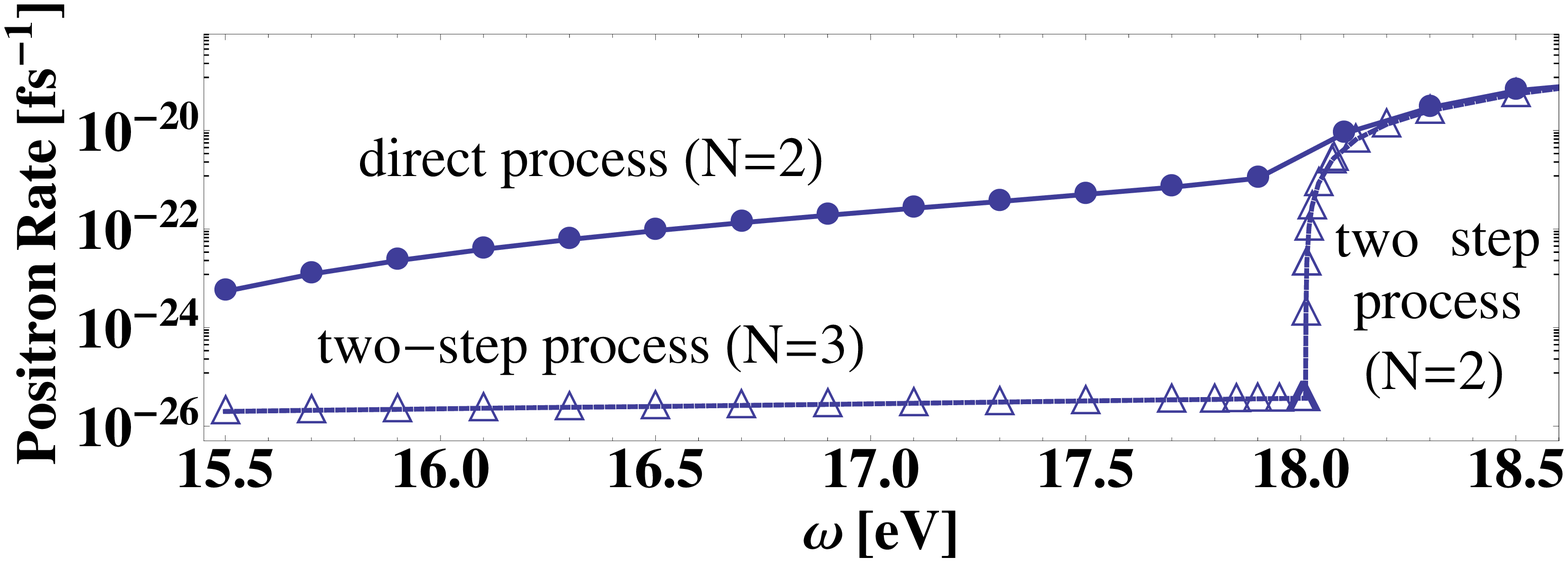}
\vspace{-0.3cm}
\caption{\label{nonres}  Laser frequency dependence of the pair creation rate in the head-on collision of a VUV pulse ($\xi=10^{-4}$) with a 17.5\,GeV electron. Shown are the separate contributions from the direct (circles) and two-step (triangles) processes whose sum yields the total rate.}
\end{figure}

In the SLAC experiment, the pair production was found to be largely dominated by the two-step mechanism \cite{SLAC}, leading to a process rate proportional to $T$ \cite{Ritus,uncertainty}. We note that the direct process (\ref{trident}) could be revealed as well by another suitable choice of collision parameters. In the laboratory frame, the threshold for the direct process with $N$-photon absorption is found to be $\omega_{\rm BH}\gtrsim\frac{2}{N}\frac{m_*}{\gamma}$, which lies below the corresponding two-step mechanism threshold
$\omega_{\rm BW}\gtrsim\frac{1}{2(\sqrt{N}-1)}\frac{m_*}{\gamma}$, with $\gamma$ being the projectile Lorentz factor. An example is shown in Fig.\,\ref{nonres}, where a 17.5 GeV electron \cite{DESY} collides with an intense soft VUV pulse \cite{VUV}. Below $\omega_{\rm BW}\approx 18\,$eV, two-photon pair creation is possible via the direct mechanism only, which can be measured separately at VUV intensities $\sim 10^{13}$\,W/cm$^2$ ($\xi=10^{-4}$). In this region, the two-step mechanism requiring an additional photon is  strongly suppressed.

The SLAC experiment relied on the high-energy electron beam from a large-scale linear accelerator. Nowadays, corresponding pair creation studies could be performed with compact laser wakefield accelerators producing few-GeV electron beams \cite{wakefield}. Assuming a laser-accelerated 5 (2) GeV electron colliding with a second optical laser pulse of intensity $\sim 10^{20}$ ($10^{21}$)\,W/cm$^2$, an observable pair creation rate of $\sim 10^5$\,s$^{-1}$ in the manifestly nonperturbative regime $\xi\approx3$ (8) results \cite{Jena}. At the envisaged high-power ($\sim 10^{25}$\,W/cm$^2$, $\xi\sim 10^3$) facility ELI \cite{ELI}, comparable rates can be expected for $p^0\sim10$\,MeV already. The 
pair creation occurs in a quasi-static regime \cite{tunneling} here since, in the electron frame, the low-frequency laser field resembles a constant crossed field \cite{Ritus} on the time scale necessary for the field to create a pair. Another all-optical scheme for pair creation in two laser beams employs a seed electron being accelerated directly by the fields \cite{Kirk}.

\begin{figure}
\includegraphics[height=4cm,width=7cm]{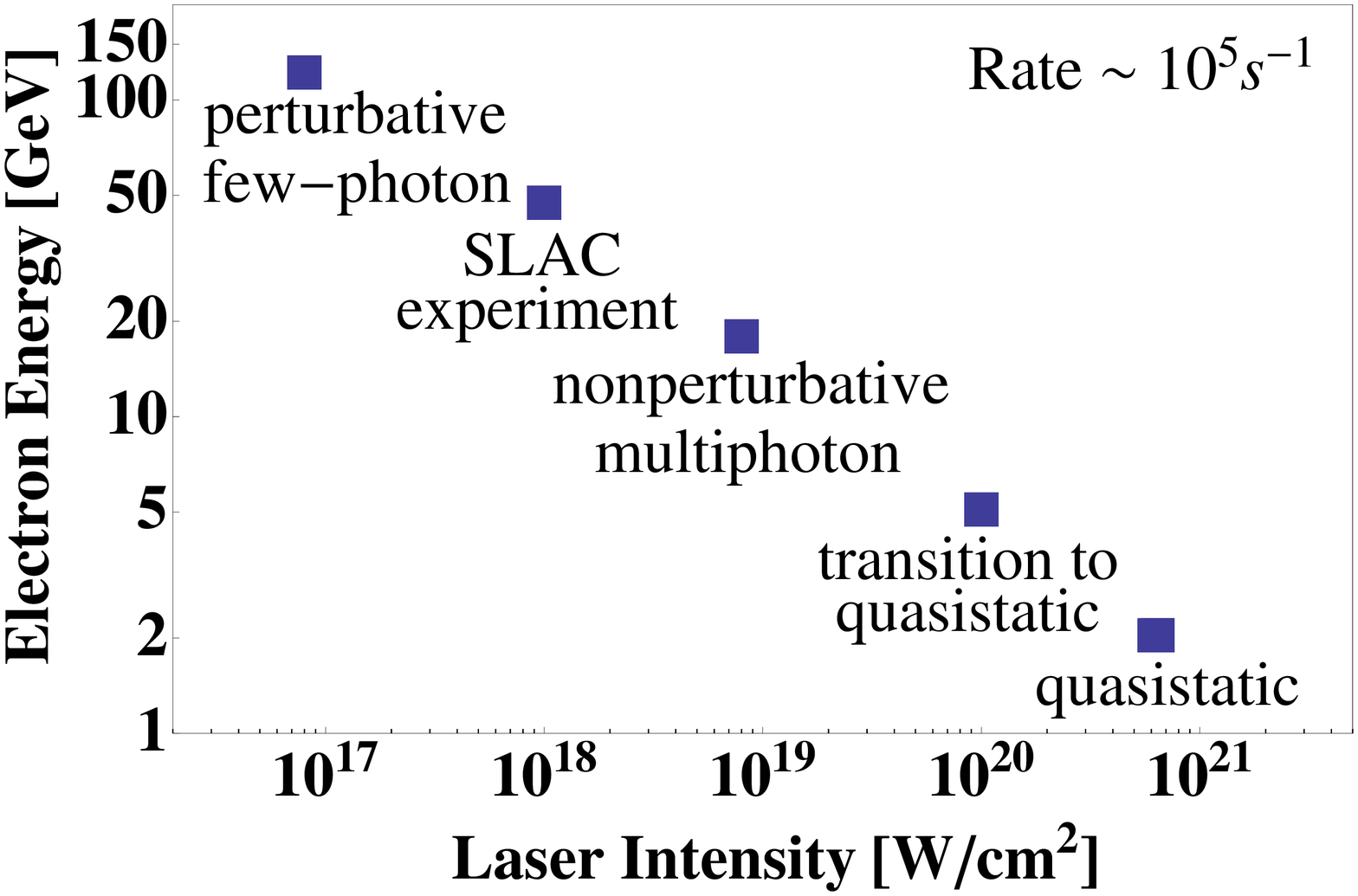}
\vspace{-0.3cm}
\caption{\label{ratefigure} Relation of the optical (527 nm) laser  intensity and the electron energy to give an observable positron rate $\sim 10^5$\,s$^{-1}$ in the lab frame. The pair creation mechanism is changing from the perturbative few-photon regime to the quasi-static (i.e. low-frequency, high-intensity) regime, as indicated.}
\end{figure}

Fig.\,\ref{ratefigure} provides an overview of our results on the total pair production rates in various interaction regimes.

In conclusion, a complete QED treatment of multiphoton trident pair creation in electron-laser collisions has been presented. It was shown that the  SLAC experiment \cite{SLAC} observed nonperturbative signatures. Future experimental studies, allowing for a disentanglement of the competing production processes in the perturbative regime and probing the transition to the fully nonperturbative domain, may rely on relativistic electrons from XFEL beamlines or laser accelerators.

We thank E. L\"otstedt, H. Bauke, H. Gies, and A. I. Milstein for useful conversations.
H.~H. acknowledges support from the National Basic Research Program of China 973 Program under Grant No. 2007CB815105 and the China Scholarship Council.


\begin{thebibliography}{99}
\bibitem{Reiss} H. R. Reiss, J. Math. Phys. \textbf{3}, 59 (1962);
Phys. Rev. Lett. \textbf{26}, 1072 (1971).

\bibitem{Nikishov} A. I. Nikishov and V. I. Ritus, Sov. Phys. JETP {\bf 19}, 529 (1964).

\bibitem{SLAC} D. Burke {\it et al.}, Phys. Rev. Lett. {\bf 79}, 1626 (1997);
C. Bamber {\it et al.}, Phys. Rev. D {\bf 60}, 092004 (1999).

\bibitem{Report} Y.\,I. Salamin {\it et al.}, Phys. Rep. {\bf 427}, 41 (2006);
M. Marklund and P.\,K. Shukla, Rev. Mod. Phys. {\bf 78}, 591 (2006).

\bibitem{Ehlotzky}
F.~Ehlotzky, K.~Krajewska, and J.~Z.~Kami\'nski, Rep. Prog. Phys. {\bf 72}, 046401 (2009).

\bibitem{MVG}
C. M\"uller, A. B. Voitkiv, and N. Gr\"un, Phys. Rev. Lett. \textbf{91}, 223601 (2003).

\bibitem{Milstein}
A. I. Milstein {\it et al.}, Phys. Rev. A \textbf{73}, 062106 (2006).

\bibitem{Kaminski} J. Z. Kami\'nski, K. Krajewska, and F. Ehlotzky, Phys. Rev. A {\bf 74}, 033402 (2006).

\bibitem{Kuchiev} M. Yu. Kuchiev, Phys. Rev. Lett. \textbf{99}, 130404 (2007).

\bibitem{Deneke}
C. M\"uller, C. Deneke, and C. H. Keitel, Phys. Rev. Lett. \textbf{101}, 060402 (2008).

\bibitem{ADP} A.~Di~Piazza {\it et al.}, Phys. Rev. Lett. \textbf{103}, 170403 (2009).

\bibitem{Ringwald} A. Ringwald, Phys. Lett. B {\bf 510}, 107 (2001).

\bibitem{Alkofer} R.~Alkofer \textit{et al.}, Phys. Rev. Lett. {\bf 87}, 193902 (2001).

\bibitem{Narozhny} N. B. Narozhny {\it et al.}, JETP Lett. \textbf{80}, 382 (2004).

\bibitem{Kirk} A.\,Bell\,and\,J.\,Kirk, Phys.\,Rev.\,Lett. \textbf{101}, 200403 (2008).

\bibitem{Gies}
F.\,Hebenstreit\,{\it et al.}, Phys.\,Rev.\,Lett. \textbf{102}, 150404 (2009).

\bibitem{Grobe} T. Cheng, Q. Su, and R. Grobe, Europhys. Lett. \textbf{86}, 13001 (2009).

\bibitem{Chen} H. Chen {\it et al.}, Phys. Rev. Lett. \textbf{102}, 105001 (2009).

\bibitem{Breit} G. Breit and J. A. Wheeler, Phys. Rev. \textbf{46}, 1087 (1934).

\bibitem{internal1} C. Bula and K. T. McDonald, E-144 Internal Note 970114 (1997);
T. Koffas and A. C. Melissinos, E-144 Internal Note 980410 (1998)
[see http:// www.slac.stanford.edu/exp/e144/notes/notes.html].

\bibitem{Volkov} D. M. Volkov, Z. Phys. {\bf 94}, 250 (1935).

\bibitem{trident} K. J. Mork, Phys. Rev. {\bf 160}, 1065 (1967).

\bibitem{Ritus} V. I. Ritus, Nucl. Phys. B \textbf{44}, 236 (1972).

\bibitem{nomenclature}
Note that in \cite{SLAC} the notion of `trident' process referred solely to the direct reaction (\ref{trident}).

\bibitem{Decay}
Under different circumstances other regulators such as the Compton decay time $1/\Gamma_C$ are relevant \cite{Ehlotzky}. In the SLAC case, though, $\Gamma_C\sim \alpha\xi^2\omega$ so that $\Gamma_C^{-1}\sim 1\,{\rm ps}\gg T_0$.

\bibitem{Reiss2} H. R. Reiss, Eur. Phys. J. D \textbf{55}, 365 (2009).

\bibitem{DESY} For current information see http://xfel.desy.de

\bibitem{uncertainty} Note that for finite interaction times, a strict distinction between off-shell and on-shell intermediate photons generally is rendered difficult by the associated energy uncertainty $\Delta {k'}^0$.

\bibitem{VUV} D. Charalambidis {\it et al.}, New J. Phys. \textbf{10}, 025018 (2008).

\bibitem{wakefield} W. P. Leemans {\it et al.}, Nature Phys. \textbf{2}, 696 (2006).

\bibitem{Jena} In a similar setup, Thomson scattering was observed 
[H. Schwoerer {\it et al.}, Phys. Rev. Lett. \textbf{96}, 014802 (2006)].

\bibitem{ELI} 
See the ELI proposal on http://www.eli-laser.eu

\bibitem{tunneling} This regime is often referred to as tunneling. See, however, H. R. Reiss, Phys. Rev. Lett. \textbf{101}, 043002 (2008), where limitations of the traditional tunneling picture in strong-field physics are identified.

\end{thebibliography}
\end{document}